\documentclass{article}
\usepackage[a4paper,
            left=1.8cm,
            right=1.8cm,
            top=2.0cm,
            bottom=2.0cm]{geometry}
\usepackage{graphicx} 
\usepackage{float}
\usepackage[normalem]{ulem}
\usepackage{xcolor} 
\usepackage{bm}          
\usepackage{amssymb}
\usepackage{amsmath}
\usepackage[backend=biber, style=ieee, isbn=false, sortcites, maxbibnames=6, minbibnames=6]{biblatex} 
\usepackage{hyperref}
\usepackage{multirow}
\usepackage{booktabs}
\usepackage{siunitx}
\newcommand{\BEQ}{\begin{equation}}     
\newcommand{\BEA}{\begin{eqnarray}}
\newcommand{\BD}{\begin{displaymath}}
\newcommand{\EEQ}{\end{equation}}       
\newcommand{\EEA}{\end{eqnarray}}
\newcommand{\ED}{\end{displaymath}}

\newcommand{\D}{{\rm d}}                

\definecolor{gruen}{rgb}{0,0.625,0}       
\definecolor{rot}{rgb}{0.75,0,0}          
\definecolor{blau}{rgb}{0,0,0.75}         
\definecolor{casta}{rgb}{0.45,0.20,0}     
\definecolor{gelb}{rgb}{0.825,0.725,0.0}  

\newcommand{\BLAU}[1]{\textcolor{black}{{\rm #1}}}	

\hypersetup{colorlinks=true,
	linkcolor=blau,
	urlcolor=blau,
    citecolor=blau}

\begin{document}


\begin{titlepage}

\vskip 1.5 cm
\begin{center}
{\LARGE \bf Fractal and Spectral Dimensions as Determinants of Thermal Ablation Outcomes in Cancer Tissues}
\end{center}

\vspace{2.0cm}
\centerline{{\bf Mario Olmo-Fajardo}$^{a,b}$, {\bf Alexander L\'opez}$^{c,d}$, {\bf Malte Henkel}$^{b,e}$, {\bf S\'ebastien Fumeron}$^{b,*}$}
\vspace{0.5cm}
\centerline{$^a$ Departamento de F\'isica, Universidad Carlos III de Madrid, E-28911 Legan\'es, Spain}
\vspace{0.5cm}
\centerline{$^b$ Laboratoire de Physique et Chimie Th\'eoriques (CNRS UMR 7019),}  
\centerline{Universit\'e de Lorraine Nancy,
B.P. 70239, F-54506 Vand{\oe}uvre l\`es Nancy Cedex, France}
 \vspace{0.5cm}
\centerline{$^c$ Escuela Superior Polit\'ecnica del Litoral, ESPOL,} 
\centerline{Departamento de F\'isica, Facultad de Ciencias Naturales y Matem\'aticas,} 
\centerline{Campus Gustavo Galindo
 Km. 30.5 V\'ia Perimetral, P. O. Box 09-01-5863, Guayaquil, Ecuador}
\vspace{0.5cm}
\centerline{$^d$ GISC, Departamento de F\'{\i}sica de Materiales, Universidad Complutense, E-28040 Madrid, Spain} 
\vspace{0.5cm}
\centerline{$^e$ Centro de F\'{i}sica Te\'{o}rica e Computacional, Universidade de Lisboa,}
\centerline{Campo Grande, P-1749-016 Lisboa, Portugal}

\let\thefootnote\relax\footnotetext{$^*$ Corresponding author. Email address: \href{mailto:sebastien.fumeron@univ-lorraine.fr}{sebastien.fumeron@univ-lorraine.fr} (S. Fumeron)}

\vspace{0.5cm}

\begin{abstract}
Clinical thermal ablation outcomes display significant variability that classical bio-heat models cannot fully explain. 
One reason may lie in the fractal architecture of biological tissues, which has been identified as a robust biomarker directly correlated with cancer grades. 
This structural heterogeneity, together with memory effects (e.g., thermotolerance), causes heat transfer in living tissues to differ from Fourier diffusion, resulting in anomalous biological transport. 
In this work, we implemented a realistic fractal–fractional bio-heat model, with non-linear perfusion and PI-controlled power delivery, 
to quantify the role of tissue fractality in ablation outcomes. 
Our results reveal that the expansion of coagulation zones is jointly controlled by fractal geometry and its associated topological connectivity. 
These findings highlight spectral dimension as a key driver of clinical variability, successfully reproducing the reduced ablative efficacy in liver metastases compared to primary carcinomas, 
and provide evidence for topologically informed treatment strategies for the thermal ablation of malignant neoplasms.

\vspace{1.5cm}

\noindent \textbf{Keywords}: Spectral dimension, Fractal dimension, Thermal ablation, Bioheat transfer, Fractional transport, Anomalous diffusion
\end{abstract}

\end{titlepage}

\setcounter{footnote}{0}

\section{Introduction}
\label{sec:intro}

Despite extensive research efforts in oncology, cancer remains one of the leading causes of death worldwide. According to the WHO~\cite{ferlay2024}, the number of new cancer cases worldwide per year in 2050 is expected to reach 35.3 million, while the number of cancer-related deaths should rise to 18.5 million. Moreover, although incremental gains in early detection and targeted therapy have greatly contributed to advancing the understanding of several cancer-related phenomena, the 5-year net survival for the major solid tumours remains stubbornly low—rarely exceeding 65\% and dropping below 20\% for the most aggressive ones~\cite{arnold2019}—underscoring the urgent need for synergetically grounded treatments that can exploit specific chemical, mechanical and physical properties of tumours.

Cancer results from a multi-scale collection of disorders that enable cells to escape the signalling pathways that regulate proliferation, differentiation, and death. At a microscopic scale, DNA mutations come from quantum mechanisms: due to a double proton tunnelling, canonical base pairs are transformed into tautomeric base pairs that can decoy DNA proofreading (L\"owdin's hypothesis)~\cite{slocombe2021quantum}. At larger scales, the expression of mutated genes and ultimately oncogenesis results from epigenetic mechanisms: these latter influence the paths (Waddington's ``chreods") followed by cells in an epigenetic curved landscape with many hills, saddle-points, and attractors, some of the latter corresponding to cancer phenotypes~\cite{davila2015modeling,aranda2018landscaping}. 

In a series of remarkable articles~\cite{hanahan2000hallmarks,hanahan2011hallmarks,hanahan2022hallmarks}, 
Hanahan and Weinberg exposed the canonical hallmarks of cancer (sustained proliferative signalling, evasion of growth suppressors, resistance to cell death, replicative immortality, angiogenesis, invasion and metastasis, 
deregulated energetics, and immune escape), providing the wanted poster of malignant tumours, but did not encode their emergent biophysical traits, such as resistance to \textit{in vivo} therapy and pleomorphism. These characteristics are crucial in many emerging therapies, first and foremost hyperthermia~\cite{Kok2020}, which consists of heating tumour cells beyond the cytotoxic threshold. Mild hyperthermia ($39\text{-}\qty{45}{\celsius}$) is designed to trigger apoptosis and sensitisation to the cytotoxic effects of chemotherapy (enhanced permeability and retention effect) and radiotherapy. Ablative hyperthermia ($> \qty{50}{\celsius}$) directly leads to tumour necrosis. Several parameters rule hyperthermia: the thermal dosimetry (setup temperature, duration, threshold effects), the kinetics of heat application (steep vs gradual heating) and the region of exposure (global, locoregional, or local).

However, rigorous control of these parameters does not guarantee consistent clinical outcomes. Recent multicentric studies~\cite{Montgomery2004, Heerink2018, Paolucci2022, Mathy2024} reveal substantial unpredictability in lesion dimensions, with applied energy explaining as little as $25\%$ of the variability in effective ablation volume~\cite{Paolucci2022}. Furthermore, the resulting necrosis frequently deviates from manufacturer specifications, often yielding significantly smaller zones and leading to aggressive over-treatment strategies to mitigate local recurrence. While the specific determinants of this variability remain poorly understood and sometimes inconsistent across the literature, the biological identity and anatomical location of the target appear to play a key role. For example, some cohorts~\cite{Heerink2018, Mathy2024} demonstrate that the ablation volumes for Hepatocellular Carcinoma (HCC) differ significantly from those of metastases, suggesting a dependency on histological origin, while lesions in perivascular locations exhibit marked geometric deviations due to the heat-sink effect. Taken together, these dependencies and the unresolved role of intrinsic physical tissue properties in ablation expansion~\cite{Paolucci2022} suggest that ablation outcomes are not governed by simple clinical descriptors alone, but rather emerge from a more complex interaction within the tissue’s physical and structural organisation.

Among the potential drivers of this clinical variability, the fractal nature of biological tissue offers a particularly compelling explanation. Extensive experimental studies have established that biological tissues exhibit statistical self-similarity, identifying the fractal dimension ($D_f$) as a robust biomarker for carcinogenesis~\cite{cross1997fractals,Tambasco2008,Lennon2015,guz2015towards,Elkington2022}. In particular, research consistently demonstrates that $D_f$ increases with tumoural evolution, registering higher values in malignant tumours compared to healthy parenchyma and scaling with disease progression. Complementing this spatial complexity, temporal heat transfer in biological media also deviates from instantaneous Fourier diffusion. Tissues exhibit significant thermal inertia and memory effects, arising not only from the finite speed of heat propagation in heterogeneous structures~\cite{Liu1996} but also from dynamic biological responses such as thermotolerance~\cite{Sapareto1978, khoei2004role}. These combined spatial and temporal irregularities result in anomalous heat diffusion that standard models cannot accurately reproduce. 
To address this, we adopt the theoretical fractal-fractional bio-heat framework established by Fumeron et al.~\cite{fumeron2025}, enhanced with non-linear physiological perfusion and a PI-controlled energy source to reflect state-of-the-art clinical scenarios.

Accordingly, the primary objective of this study is to implement this comprehensive model numerically to quantify the impact of tissue fractality 
on thermal dosimetry and the resulting ablation zones across a broad spectrum of anomalous transport regimes driven by non-Markovian dynamics. In particular, we reproduce a clinically observed anomaly—the significantly lower ablative efficacy in liver metastases (LM) compared with primary hepatocellular carcinomas (HCC)—thereby enabling a physical interpretation of the spectral dimension ($d_s$)~\cite{alexander1982} in biological tissue. The remainder of this paper is organised as follows: Section~\ref{sec:model} details the mathematical formulation—integrating the fractal Pennes equation and memory effects with non-linear blood perfusion and cell viability models—as well as the thermal energy delivery protocol and numerical framework. Section~\ref{sec:results} explores the spatio-temporal dynamics of the ablation front, quantifying the impact of topological uncertainty and applying this theoretical framework to unravel the origins of clinical variability between primary and metastatic tumours. Finally, Section~\ref{sec:conclusions} summarises the main conclusions and outlines future perspectives.

\section{The model}
\label{sec:model}

\subsection{Fractal Pennes-based bio-heat model with memory effect}
 Since its formulation in 1948~\cite{pennes1948}, the Pennes bio-heat equation has become the standard baseline model for analytical and numerical research on bio-heat transfer. 
 It describes the temporal evolution of the temperature field in biological tissue by establishing a balance between Fourier's conductive heat loss (diffusion), 
 convective heat loss through the bloodstream (blood perfusion), metabolic generation of heat, and heat delivered by an external source. It takes the following form:
\begin{equation}
    \rho_t c_t \frac{\partial T}{\partial t} = \nabla \cdot (\lambda \nabla T) - \rho_b c_b \omega_b (T - T_b) + Q_m + Q_s,
    \label{eq:pennes}
\end{equation}
where $T$ is the temperature of the tissue; $t$ is the time; $\rho_t$ and $c_t$ are the density and specific heat of the tissue, respectively; $\lambda$ is the thermal conductivity; $\rho_b$ and $c_b$ are, respectively, the density and specific heat of the blood; $\omega_b$ is the blood perfusion rate; $T_b$ is the arterial blood temperature; $Q_m$ is the metabolic heat generation term; and $Q_s$ represents the external heat source.

Recently, a new Pennes-based fractal bio-heat model has been proposed~\cite{fumeron2025}, which incorporates two fundamental features of neoplastic tissue that were missing in the original formulation: the non-locality in time (thermotolerance) and the fractal structure of tissue (pleomorphism). Tumours can sustain hyperthermia more effectively than normal cells as a result of two mechanisms: intrinsic thermoresistance and acquired thermotolerance. Thermoresistance refers to the ability of tumours to survive higher temperatures compared to normal cells—for example by failing to activate caspase-3~\cite{tang2020cancer}—which limits the cytotoxicity of thermal therapies. In addition, tumours can develop thermotolerance: this is a rapidly developing phenomenon that peaks within 24 hours of treatment and decays slowly over 3 to 5 days, depending on the dose rather than on the temperature level~\cite{urano1986kinetics}. The molecular origin of thermotolerance is the up-regulation of heat shock proteins (such as HSP70, HSP90 and small HSPs) that act as molecular chaperones, folding denatured proteins and preventing protein aggregation~\cite{khoei2004role}. In prostate cancer, the activity of HSPs, and thus thermotolerance, can be inhibited pharmacologically with quercetin. While these molecular mechanisms operate on longer timescales (days), tissues also exhibit immediate thermal inertia—arising from non-Fourier finite-speed heat propagation and thermal wave mechanisms~\cite{Liu1996}—and transient structural adaptation during acute heating~\cite{Welch1985}. To phenomenologically capture the non-local memory effects inherent to these dynamic responses we employ a temporal Caputo fractional derivative of non-integer order $\alpha$ ($n-1\leq \alpha < n$, $n\in \mathbb{N}$):
\begin{equation}
    \label{eq:CaputoMod}
    \frac{\partial T(t)}{\partial t} 
    \longrightarrow \tau^{\alpha-1} \,{}^{C}\!D_{t}^{\alpha} T(t) = \frac{\tau^{\alpha-1}}{\Gamma(n - \alpha)} \int_{0}^t (t - u)^{ n -\alpha - 1} \frac{\D^nT(u)}{\D u^n} \, \D u,
\end{equation}
where the relaxation time $\tau$ ensures dimensional consistency and scales the heat propagation velocity. \BLAU{Among the existing non-local operators (e.g., the Riemann-Liouville derivative), the Caputo definition is chosen given its suitability for initial-value problems. Since the Caputo derivative of a constant is zero, it allows the formulation of standard integer-order initial conditions—such as the initial tissue temperature and its rate of change—that preserve a clear physical interpretation~\cite{diethelm_2010}.} Depending on the fractional order, the tissue temperature response exhibits three distinct bio-heat transfer regimes~\cite{fumeron2025}: a damped (sub-diffusive) regime for $0<\alpha<1$, a critically damped super-diffusive regime for $1<\alpha\leq1.5$, and an underdamped oscillatory super-diffusive regime for $1.5<\alpha<2$. Classical diffusion is recovered when $\alpha=1$.

In addition, similar to many other ubiquitous natural structures (such as coastlines, snowflakes, galaxy clusters, or turbulent flows), biological tissue 
\BLAU{can be described as a natural fractal}. Fractals, as defined by Mandelbrot~\cite{mandelbrot1983}, are mathematical structures that exhibit self-similarity—where each part is geometrically similar to the whole—across multiple scales. 
\BLAU{In the case of biological tissues, this self-similarity is statistical rather than deterministic. Furthermore, unlike these ideal mathematical constructs, natural physical structures inherently possess a finite range of self-similarity bounded by physical lower and upper cut-offs (e.g., from cellular dimensions up to macroscopic tissue sizes).} Characterisation of a fractal typically requires at least three parameters~\cite{rammal1983}: the embedding Euclidean dimension $d$, the fractal dimension $D_f$, and the spectral dimension $d_s$. These parameters satisfy the inequality $d_s \leq D_f \leq d$, which reduces to $d_s = D_f = d$ in regular Euclidean spaces. The fractal dimension, which is related to the geometry and space-filling capacity of the structure, is a well-studied property of biological tissue. It can be determined experimentally from microscopic imaging of histological cuts and serves as a morphometric biomarker in oncopathology~\cite{cross1997fractals,Tambasco2008, Lennon2015, guz2015towards,Elkington2022}. The clinical range extends from $D_f \approx 1.5$ for healthy tissue to $D_f \approx 1.8$ for advanced-stage tumours, showing an upward trend as the disease evolves. In contrast, the spectral dimension $d_s$, which governs the topology and connectivity of the fractal, remains unexplored in biological tissue. Despite previous efforts to determine $d_s$ in percolation clusters and other mathematical fractals~\cite{alexander1982}, there is currently no method for measuring the spectral dimension in biological fractals. Diffusion in such media is anomalous, and consequently, the Laplace operator must be modified according to:
\begin{equation}
    \label{eq:FractalsMod}
    \nabla \cdot (\lambda \nabla T) = 
    \frac{1}{r^{d-1}}\frac{\partial}{\partial r}\left( \lambda r^{d-1}\frac{\partial T}{\partial r} \right) 
    \longrightarrow \frac{1}{r^{D_f-1}}\frac{\partial}{\partial r}\left( \tilde{\lambda}r^{-\theta} r^{D_f-1}\frac{\partial T}{\partial r} \right),
\end{equation}
where $\tilde{\lambda}$ has units of~\si{\watt\meter^{\theta-1}\kelvin^{-1}} and the exponent $\theta = (2D_f/d_s) - 2 > 0$ 
accounts for anomalous diffusion, characterised by a mean squared displacement scaling as $\langle r^2 \rangle \propto t^{2/(2+\theta)}$~\cite{alexander1982, OShaughnessy1984, OShaughnessy1985}. 
The generalised units of $\tilde{\lambda}$ hinder the application of the previous expressions to real clinical settings, in which the conductivity is typically measured in SI units. 
For this reason, a characteristic length $\ell$ (\si{\meter}) is introduced, allowing the definition of the effective conductivity $\lambda_{\rm eff}(r) = \lambda_0 (r/\ell)^{-\theta}$, 
which is expressed in terms of the baseline conductivity $\lambda_0$ (\si{\watt\meter^{-1}\kelvin^{-1}}) reported in the literature. 
While $\ell$ may look like a simple scaling constant, it determines the physics of transport: for $\ell < r$ the effective conductivity decreases as $\theta$ increases 
(sub-diffusion), whereas for $\ell > r$ the opposite behaviour arises. 
Considering that diffusion in fractals is sub-diffusive—the second moment of the displacement decreases for increasing values of $\theta$—the case in which 
$\ell < r$ is the only physically admissible scenario. 
Connecting this fact with the notion of 
\BLAU{a natural} fractal, this characteristic length $\ell$ corresponds to the 
\BLAU{physical lower cut-off from which} the self-similarity property holds; 
in other words, the dimensions of the minimum constituent of the fractal. 
In this work, this characteristic length is assumed to be equal to the microscopic scale at which the fractal dimension of biological tissue has been measured in the literature 
($\sim 100$~\si{\micro\meter})~\cite{Lennon2015, Elkington2022}. 
This implies that diffusion at lower scales is ruled by classical Fickian dynamics, which are not captured by the previous fractal formulation alone. 
Thus, the super-diffusive regime emerging when $\ell > r$ is numerically avoided by imposing a spatial step size $\Delta r \geq \ell$.

Combining these modifications, the resulting 2D fractal bio-heat model with memory effects can be written as:
\begin{equation}
    \label{eq:ModPennes}
    \rho_t c_t \tau^{\alpha-1} \, {}^{C}\!D_{t}^{\alpha}T 
    = \frac{1}{r^{D_f-1}}\frac{\partial}{\partial r}\left( \lambda_0\left(\frac{r}{\ell}\right)^{-\theta} r^{D_f-1}\frac{\partial T}{\partial r} \right) - \rho_b c_b \omega_b (T - T_b) + Q_m + Q_s.
\end{equation}

\subsection{Non-linear blood perfusion model accounting for vasodilation}
Blood perfusion plays a key thermoregulatory role in tissues during hyperthermia treatments. 
Several experimental studies have reported a marked increase in microvascular perfusion at $41\text{-}\qty{45}{\celsius}$ due to vasodilation, 
followed by a decrease attributed to vascular collapse resulting from thermal damage~\cite{Dewhirst1984, Xu1998, He2004}. 
This dynamic vascular response critically shapes lesion size, 
causing bio-heat models with constant blood perfusion to underestimate the ablation zone and risk damage to adjacent healthy tissue~\cite{Soni2015}.

This phenomenon can be described by the degree of stasis (DS), a parameter reflecting the fractional injury to the tumour vasculature~\cite{He2004}.

\begin{equation}
    \label{eq:DS}
    DS(r,t) = 1-\exp\left(-\xi(r,t)\right),
\end{equation}
where $\xi(r,t)$ captures microvascular thermal damage through a first-order Arrhenius kinetic model:
\begin{equation}
    \label{eq:TD_perf}
    \xi(r,t) = \int_0^t A_{\rm perf}\exp\left(-\frac{\Delta E_{\rm perf}}{RT(r,\tau)}\right) \, \D\tau,
\end{equation}
where $A_{\rm perf}$ is the frequency factor, $\Delta E_{\rm perf}$ is the activation energy of the vascular lesion process, $R$ is the universal gas constant and $T$ is the local tissue temperature.

A relation between DS and the blood perfusion parameter, $\omega_b$, was determined experimentally in porcine kidney tissue~\cite{He2004} 
and modelled as a piecewise function~\cite{Schutt2008} as follows:
\begin{equation}
    \label{eq:omega}
    \omega_b(DS) =
    \begin{cases}
        \omega_{b,0} \left(1 + 30\cdot DS\right), & DS \leq 0.02 \\
        \omega_{b,0} \left(1.86 - 13\cdot DS\right), & 0.02 < DS \leq 0.08 \\
        \omega_{b,0} \left(0.884 - 0.79\cdot DS\right), & 0.08 < DS \leq 0.97 \\
        \omega_{b,0} \left(3.87 - 3.87\cdot DS\right), & 0.97 < DS \leq 1
    \end{cases}
\end{equation}
where $\omega_{b,0}$ is the baseline blood perfusion parameter.
Recent studies on ultrasound thermal ablation~\cite{Prakash2012} and nanoparticle-assisted thermal therapy~\cite{Soni2015, Singh2022} 
have employed the non-linear blood perfusion model in combination with the classical Pennes bio-heat equation.

\subsection{Assessment of thermal damage and cell viability}
Two models are commonly used to evaluate thermal injury in biological tissue: 
the first-order Arrhenius kinetic damage model~\cite{Henriques1947a, Moritz1947a, Moritz1947b, Henriques1947b} and the cumulative equivalent minutes at \qty{43}{\celsius} (CEM43) model~\cite{Sapareto1984}. 
Both approaches account for the cumulative effect of temperature over time, since the extent of thermal damage is determined by the temperature-time history of the tissue.

In this study, thermal injury is assessed using the Arrhenius formulation, 
as it provides a better estimation of damage at high temperatures and allows a direct physical interpretation through its relation to the viable cell concentration~\cite{Pearce2013}. 
Thus, thermal damage ($\Omega$) is given by:
\begin{equation}
    \label{eq:TD_reg}
    \Omega(r,t) = \int_0^t A\exp\left(-\frac{\Delta E}{RT(r,\tau)}\right) \, \D\tau,
\end{equation}
and can also be recast as 
\begin{equation}
    \label{eq:SR}
    \Omega(r,t) = -\ln \frac{c(t)}{c(t_0)},
\end{equation}
where $A$ is the frequency factor, $\Delta E$ is the activation energy of the tissue lesion process, $R$ is the universal gas constant, and $T$ is the local tissue temperature. 
The terms $c(t_0)$ and $c(t)$ represent the initial and remaining viable cell concentrations, respectively.

In this sense, the {\em coagulation ablation zone} is defined as the region where the thermal damage parameter, $\Omega$, is greater than or equal to 4.6, 
corresponding to approximately 99\% cell death. Another relevant region considered in this work is the {\em periablation zone}, defined in~\cite{Trujillo2020} 
as the tissue between thresholds of $\Omega = 0.6$ ($\approx 45\%$ cell death) and $\Omega = 2.1$ ($\approx 88\%$ cell death). 
This zone surrounds the central coagulation area and contains both damaged and viable cells. 
Recent studies have shown that sublethal thermal injury experienced by cells within this region can promote distant tumour growth~\cite{Velez2016, Markezana2020, Wan2024}. 
Consequently, the extent of the periablation zone becomes a critical parameter to minimise during ablative therapies.

\subsection{Thermal energy delivery and control}
The most common heating techniques in hyperthermic and ablative treatments include electromagnetic (RF, IR, laser), ultrasound, perfusion, and conductive heating~\cite{Kok2020}. 
Regardless of the specific energy source used, two main approaches are usually followed in clinical and modelled ablation control: 
constant power (open-loop) and temperature control (closed-loop)~\cite{Berjano2006}.

Although the simplicity of the constant power mode made it the preferred approach in early ablation and some current experimental studies, 
it requires a fine tuning of the applied power level and of its duration in order to avoid unwanted tissue damage and vaporisation at $T>\qty{100}{\celsius}$. 
The latter leads to the formation of microbubbles~\cite{Wood2005} associated with a decrease in the electrical conductivity of tissue, a fundamental property in RF ablation. 
However, the effect of vaporisation on thermal conductivity was shown to be negligible~\cite{Trujillo2013}.

In contrast, temperature-controlled modes modulate the power amplitude in real time to maintain a target temperature at a specific sensor location, ensuring safety and reproducibility. 
In addition, as first shown in~\cite{Jain1999}, they drive the temperature profile to a steady-state which eventually limits the coagulation ablation zone to a specific size. 
This temperature control is achieved through a closed-loop feedback system, typically implemented by a Proportional-Integral (PI) controller~\cite{Haemmerich2005}. 
In this regard, several clinical and theoretical studies have been conducted on the effectiveness and suitability of 
different temperature-controlled heating protocols~\cite{Zhi-yu2017, lu2019, chen2019, Rivas2021, Bottiglieri2025}.

In order to mimic the energy deposition pattern of real ablation treatments, the volumetric heat generation source, $Q_{s}$, 
is modelled in this study by using a Gaussian distribution function centred at $r=0$ and modulated by a PI controller:
\begin{equation}
    \label{eq:source}
    Q_s(r,t) = Q_{s,0} \cdot \exp \left(-\left(\frac{r}{\sigma}\right)^2\right)\cdot f_{fb}(T(0,t))\cdot g(t),
\end{equation}
where $Q_{s,0}$ is the peak volumetric heat source at the application point and $\sigma$ is its characteristic radius. The feedback modulation factor, 
$f_{fb} \in [0,1]$, determines the fraction of power applied based on the temperature error, $e(t)$, between the target limit temperature, 
$T_{\rm lim}$ and the temperature at the application point, $T(r=0, t)$:
\begin{equation}
    \label{eq:source_feedback}
    f_{fb}(T(0,t)) = \text{sat}_{[0,1]} \left(K_pe(t) + K_i\int_0^te(\tau) \, \D\tau\right),
\end{equation}
where $e(t) = T_{\rm lim} - T(0,t)$ and $K_p$ and $K_i$ are the proportional and integral gain constants, respectively. The saturation function, $\text{sat}_{[0,1]}(x)$, 
bounds the control signal to the physically realisable power output and is formally defined for any input $x$ as:
\begin{equation}
\label{eq:saturation_def}
\text{sat}_{[0,1]}(x) =
\begin{cases}
0, & x < 0 \\
x, & 0 \leq x \leq 1 \\
1, & x > 1
\end{cases}
\end{equation}
To prevent integral windup when the actuator saturates (i.e., when the required power exceeds the physical limits), a back-calculation anti-windup strategy was implemented in the numerical loop. 
Finally, the duration of the external heat generation, $t_{\rm on}$, is ruled by the binary switching function $g(t)$:
\begin{equation}
    \label{eq:source_time}
    g(t) =
    \begin{cases}
        1, & 0\leq t \leq t_{\rm on} \\
        0, & \text{otherwise}
    \end{cases}
\end{equation}

\subsection{Numerical framework}
\begin{table}[h]
    \centering
    \small
    \begin{tabular}{@{}llcc@{}}
        \toprule
        \textbf{Parameter} & \textbf{Symbol} & \textbf{Value} & \textbf{Units} \\ 
        \midrule
        \multicolumn{4}{@{}l}{\textit{Bio-heat Properties (Tissue \& Blood)}~\cite{Singh2020, Singh2022}} \\ 
        Density & $\rho_t, \, \rho_b$ & 1060 &~\si{\kilogram\per\meter\cubed} \\
        Specific heat capacity & $c_t,\,c_b$ & 3780 &~\si{\joule\per\kilogram\per\kelvin} \\
        Baseline thermal conductivity & $\lambda_0$ & 0.5 &~\si{\watt\per\meter\per\kelvin} \\
        Arterial blood temperature & $T_b$ & 37 &~\si{\celsius} \\
        Baseline blood perfusion & $\omega_{b,0}$ & $1.11 \times 10^{-3}$ &~\si{\per\second} \\
        \midrule
        \multicolumn{4}{@{}l}{\textit{Arrhenius Damage Model}~\cite{Borrelli1990, Brown1992, Sherar2000, Breen2007}} \\
        \textbf{Vascular Stasis:} & & & \\
        \hspace{3mm} Frequency factor & $A_{\rm perf}$ & $1.98\times10^{106}$ &~\si{\per\second} \\
        \hspace{3mm} Activation energy & $\Delta E_{\text{perf}}$ & $6.67\times10^{5}$ &~\si{\joule\per\mole} \\
        \textbf{Thermal Damage:} & & & \\
        \hspace{3mm} Frequency factor & $A$ & $2.98\times10^{80}$ &~\si{\per\second} \\
        \hspace{3mm} Activation energy & $\Delta E$ & $5.06\times10^{5}$ &~\si{\joule\per\mole} \\
        \midrule
        \multicolumn{4}{@{}l}{\textit{Source Term \& PI Controller}} \\
        Peak volumetric heat source& $Q_{s,0}$ & $6\times10^{6}$ &~\si{\watt\per\meter\cubed} \\
        Source radius & $\sigma$ & $3\times10^{-3}$ &~\si{\meter} \\
        Limit temperature & $T_{\rm lim}$ & 90 &~\si{\celsius} \\
        Ablation duration & $t_{\rm on}$ & 600 &~\si{\second} \\
        Proportional gain & $K_p$ & 1.3 &~\si{\per\kelvin} \\
        Integral gain & $K_i$ & 0.15 &~\si{\per\kelvin\per\second} \\ 
        \bottomrule
    \end{tabular}
    \caption{Summary of thermophysical properties, Arrhenius model parameters, and heat source controller settings employed in the numerical simulations.}
    \label{tab:simulation_parameters}
\end{table}
Numerical simulations were performed on a radially symmetric domain representing a cross-section of an idealised tumour. Under the assumption of angular independence (azimuthal isotropy), the spatial discretisation was applied along the radial direction $r$, extending from the tumour centre to the distal boundary ($R=15$~\si{\centi\meter}).

The tissue domain was initialised at a uniform physiological steady state ($T(r,0) = T_0$), with vanishing initial time derivative ($\partial T/\partial t \,(r,0) = 0$). The latter condition is strictly required for fractional order $1<\alpha\leq2$, ensuring thermodynamic equilibrium prior to ablation. Regarding boundary conditions, a symmetry Neumann condition was imposed at the tumour centre ($\partial T/\partial r \,(0,t) = 0$) and constant homeostatic temperature was fixed at the distal boundary ($T(R,t) = T_0$), sufficiently far to avoid boundary artifacts.

The numerical solution of Eq.~\ref{eq:ModPennes} consisted of a fractional Predictor-Corrector Finite Difference Method (FDM) 
scheme based on the classical second-order Adams–Bashforth–Moulton method~\cite{diethelm_2010}. 
It was implemented in Python leveraging CPU parallelisation via Numba for computational efficiency. 
A spatial discretisation of $\Delta r=0.5$~\si{\milli\meter} and a temporal step of $\Delta t=0.1$~s were employed in order to satisfy stability criteria and avoid the non-physical regime arising when 
$\ell>r$. The characteristic length scale was set to $\ell=0.1$~\si{\milli\meter}, consistent with the characteristic fractal scale reported in the literature~\cite{Lennon2015, Elkington2022}. 
The relaxation time was fixed at a reference value of $\tau=1$~\si{\second}, selected after verifying that the model yields consistent results across a span of two orders of magnitude. 
The baseline thermophysical properties for tumour tissue, along with the constitutive parameters for the Arrhenius damage models and the PI controller settings employed in the simulations, 
are presented in Table~\ref{tab:simulation_parameters}. Metabolic heat generation $Q_m$ was considered negligible. 
Finally, a parametric robustness and sensitivity analysis exploring variations in $\ell$, $\tau$, and $T_{\rm lim}$ is provided in Appendix~\ref{sec:appendix_sensitivity} (Fig.~\ref{fig:sensitivity_panel}).

To quantify the relative influence of geometrical ($D_f$) and topological ($d_s$) parameters on the ablation outcome, 
we calculated a dimensionless normalised sensitivity index ($S$)~\cite{Jing2025, Ingalls2013}. 
This index represents the local relative sensitivity of the coagulation ablation radius ($r_c$) with respect to a baseline parameter $p$, defined as:
\begin{equation}
    \label{eq:sensitivity}
    S = \left| \frac{\partial \ln r_c}{\partial \ln p} \right| = \frac{p}{r_c}\left| \frac{\partial r_c}{\partial p} \right| \approx \frac{p}{r_c} \left| \frac{r_c(p(1+\epsilon)) - r_c(p(1-\epsilon))}{2\epsilon p} \right|,
\end{equation}
where $r_c$ corresponds to the radius at the unperturbed baseline value $p$. Given the non-linear nature of the bio-heat model, the derivative was approximated using a central difference scheme to minimise directional bias. A symmetric perturbation of $\epsilon = 0.05$ (5\%) was applied ($p_{\pm} = p(1 \pm \epsilon)$) to ensure numerical robustness against discretisation noise while capturing local non-linearities.

\section{Results and discussion}
\label{sec:results}

\subsection{Anomalous transport dynamics under clinical control constraints}

\begin{figure}[htbp]
  \centering
  \includegraphics[width=\linewidth]{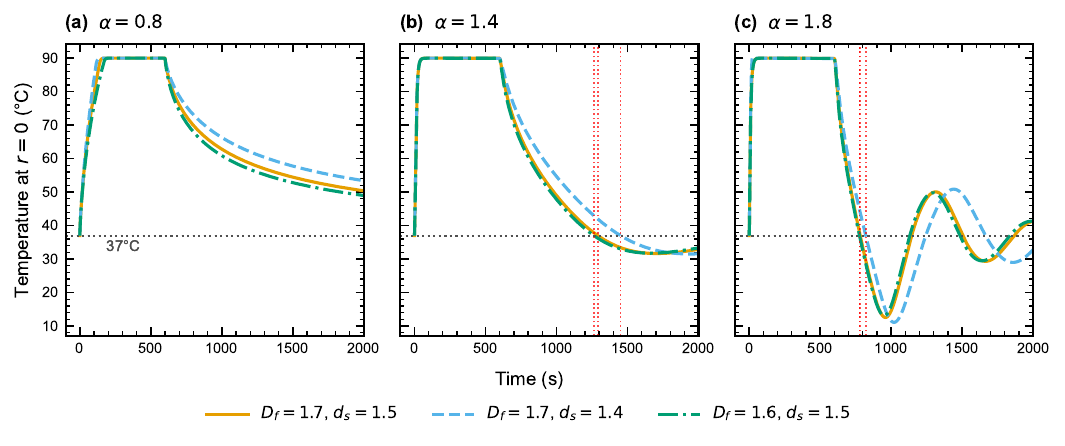}
  \caption{
    \textbf{Temporal evolution and PI control at the heating centre.} The temperature history at the application point ($r=0$) is shown for varying fractal geometries and fractional orders: (a) $\alpha=0.8$, (b) $\alpha=1.4$, and (c) $\alpha=1.8$. The feedback loop effectively constrains the maximum temperature to the target setpoint $T_{\rm lim}=\qty{90}{\celsius}$, ensuring a controlled ablative plateau. Grey horizontal lines denote the homeostatic temperature (\qty{37}{\celsius}); red vertical lines indicate the onset of unphysical cooling below this threshold.
    }
  \label{fig:T_vs_t_r0}
\end{figure}

To ensure clinical realism, the interplay between anomalous diffusion and energy delivery was evaluated under closed-loop control constraints across various fractal parameter sets ($D_f, d_s$). \BLAU{As an initial mathematical exploration, all diffusive regimes ($\alpha$) were studied, aiming to identify the physically relevant range for $\alpha$ before the comparison with clinical data.} Despite the highly non-linear nature of fractional transport, the PI algorithm successfully constrained the maximum temperature to the target setpoint ($T_{\rm lim} = \qty{90}{\celsius}$) in all cases (Fig.~\ref{fig:T_vs_t_r0}). The temporal evolution of the temperature field at the applicator tip reveals distinct dynamic behaviours depending on the fractional order $\alpha$. Notably, in the sub-diffusive case ($\alpha=0.8$), the temperature profile exhibits a markedly slow dynamic response and a trapping effect, maintaining hyperthermic temperatures ($>\qty{50}{\celsius}$) even after $\approx 23$~\si{\minute} of cooling. Conversely, the super-diffusive regimes ($\alpha=1.4$ and $\alpha=1.8$) experience rapid long-range thermal transport, causing the target temperature to drop sharply once the generator is turned off, \BLAU{reaching values even below the homeostatic temperature. This latter out-of-equilibrium oscillatory thermodynamic response is reminiscent of behaviours reported in non-equilibrium spatio-temporal nanoscale systems, where fluctuation theorems allow for apparent local violations of the second law of thermodynamics \cite{Wang2002, Gieseler2014, Paraguassu2025}, and heat transfer transitions from standard to wave-like non-Fourier thermal transport \cite{Mazza2021, Akbarzadeh2017}. Consequently, despite being particularly suitable for ultrafast heating processes, super-diffusive regimes ($\alpha > 1$) are not physically representative of macroscopic continuous thermal ablation applications, where the probability of such fluctuations is negligible and the temperature cannot descend below that of the surrounding tissue. For this reason, all subsequent clinical analyses and simulations in this study are strictly restricted to physically admissible regimes ($\alpha \le 1$).} Regarding the impact of fractal and spectral dimensions on the temperature field, the most notable difference appears as $d_s$ is decreased, hindering thermal diffusion due to decreased medium connectivity. All previous observations are consistent with the analytical regimes described in~\cite{fumeron2025}.

\begin{figure}[H]
  \centering
  \includegraphics[width=\linewidth]{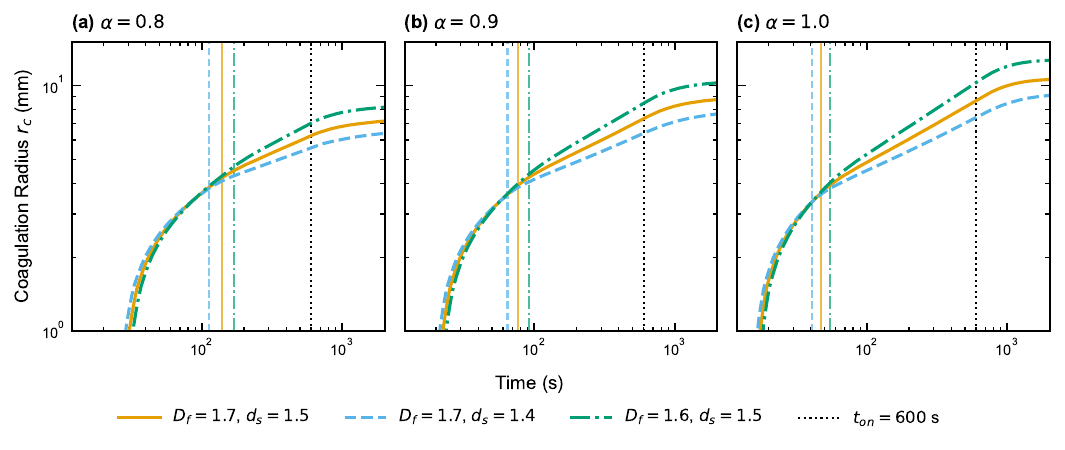}
  \caption{
    \textbf{Log-log plot of coagulation radius evolution over time.} The coagulation radius ($\Omega=4.6$) is shown for three fractional orders:
    (a) $\alpha=0.8$, (b) $\alpha=0.9$, and (c) $\alpha=1.0$. Curves in each panel correspond to different fractal parameter sets \{$D_f, d_s$\}. The black dotted line indicates the end of heating ($t_{\rm on}=600$ s). Additionally, thin vertical lines, coloured to match their corresponding curves, indicate the time at which the PI controller starts modulating the heat source.
    }
  \label{fig:r_coag_time}
\end{figure}

The translation of these anomalous transport dynamics into irreversible tissue damage is illustrated by the time-dependent expansion of the coagulation zone $r_c$ (Fig.~\ref{fig:r_coag_time}). A key observation from these logarithmic plots is that the constant-temperature control strategy—routine in clinical settings—produces an effective splitting of the coagulation radius as a function of $D_f$ and $d_s$. While the ablation front expands uniformly regardless of the tissue's fractal nature prior to the activation of the PI controller, the closed-loop power control introduces a deep dependence on the specific geometry and topology of the medium. The growth rate of this coagulation front is dictated by the fractional order $\alpha$, both before and after the action of the PI controller. As $\alpha$ decreases, so does the growth rate of the ablation radius. In addition, the expansion of the coagulation radius following the end of the heating phase is markedly reduced as $\alpha$ decreases; the trapping effect illustrated in Fig.~\ref{fig:T_vs_t_r0}a at the application point leads to hindered thermal transport during the cooling phase.

\begin{figure}[htbp]
  \centering
  \includegraphics[width=\linewidth]{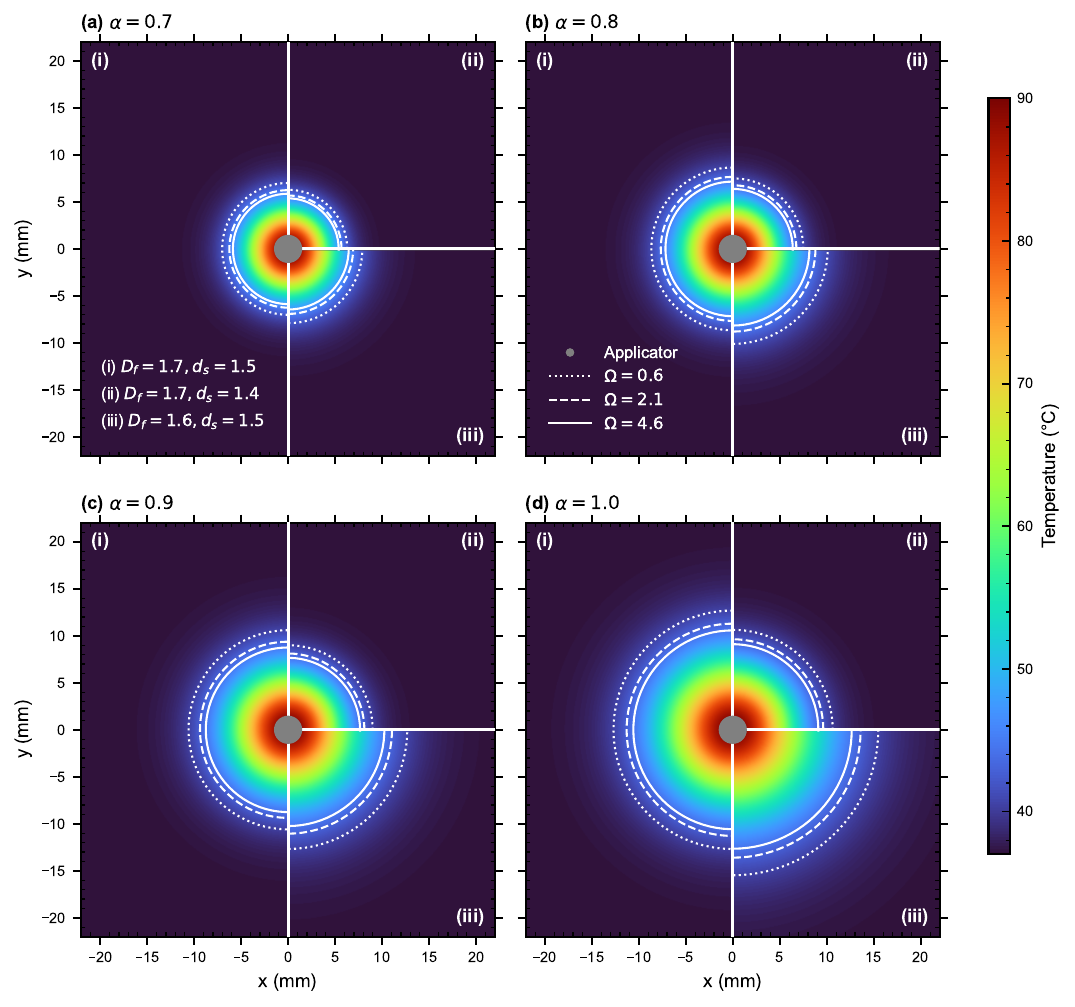}
  \caption{
    \textbf{Effect of fractional order $\alpha$ and fractal parameters ($D_f, d_s$) on spatial temperature distribution.}
    Heatmaps of temperature $T(x,y)$ at the end of the heating phase ($t_{\rm on} = 600$ s). Each panel (a-d) corresponds to a different fractional order $\alpha$, as indicated. To visualise the effect of fractal and spectral dimensions, each panel is a composite of three distinct simulations: Region (i) [left] corresponds to ($D_f=1.7, d_s=1.5$); Region (ii) [top-right] corresponds to ($D_f=1.7, d_s=1.4$); Region (iii) [bottom-right] corresponds to ($D_f=1.6, d_s=1.5$). Overlaid white lines show the simulated ablation contours for coagulation ($\Omega=4.6$, solid) and periablation ($\Omega=2.1$, dashed; $\Omega=0.6$, dotted). Contours were computed after a cooling period of $\approx23$~\si{\minute}). The central 3~\si{\milli\meter} $\emptyset$ applicator is shown in gray. Legends defining regions, contours, and the applicator are embedded in panels (a) and (b).
    }
 \label{fig:heatmap_composite}
\end{figure}

Mapping the aforementioned temporal dynamics into spatial outcomes, the radial temperature gradients and final ablation zones prove highly sensitive to the underlying tissue geometry and topology across the distinct $\alpha$-dependent regimes (Fig.~\ref{fig:heatmap_composite}). A reduction in the fractal dimension $D_f$ (and thus in structural complexity) leads to a greater thermal expansion, increasing the final size of the coagulation zone. Conversely, as the connectivity driven by the spectral dimension $d_s$ decreases, the thermal trapping effect is enhanced and the ablation radii shrink. These geometric and topological effects are independent of the fractional order $\alpha$, which modulates the spatial extent of the ablative temperature front and its associated thermal damage, with lower values of $\alpha$ leading to a reduced spatial reach. \BLAU{However, a decreased sensitivity of $r_c$ to both geometry and topology becomes evident as $\alpha$ decreases.} Notably, the size of the periablation zone—critical in thermal ablation—is proportionally greater (compared to the size of the coagulation zone) \BLAU{as $\alpha$ approaches the standard limit ($\alpha = 1$).}

\subsection{The critical role of the spectral dimension: uncertainty and sensitivity analysis}

\begin{figure}[htbp]
  \centering
  \includegraphics[width=\textwidth]{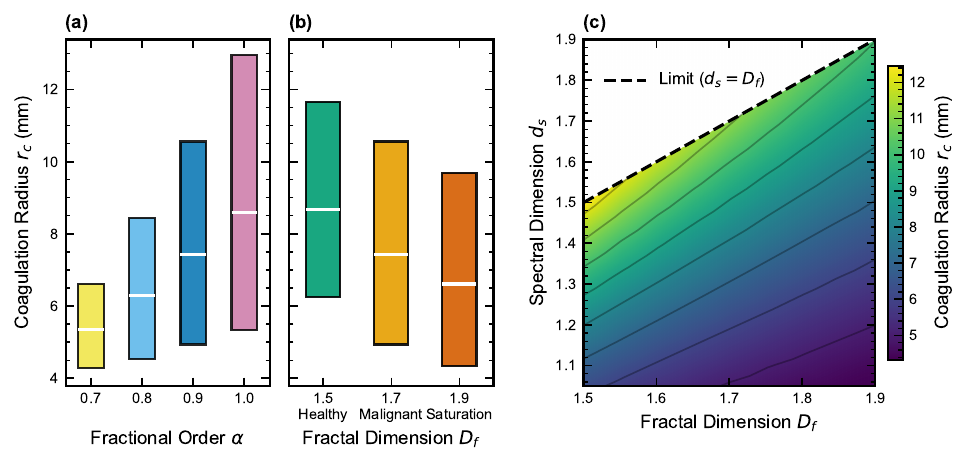}
  \caption{
  \textbf{Impact of topological uncertainty on ablation predictability.} 
  The floating bars (a,b) quantify the theoretical variability range of the final coagulation radius ($r_c$) resulting from the indeterminacy of the spectral dimension ($d_s$), 
  which is varied across a broad range of physically plausible values and strictly bounded by $D_f$ ($1.05 \leq d_s \le D_f - 0.05$). 
  The white horizontal marker indicates the response at the mid-point of the parameter range.
  (a) Variability across fractional diffusion regimes ($\alpha$) assuming a fixed malignant tumour geometry ($D_f=1.7$).
  (b) Variability across tissue fractal dimensions ($D_f$) for a fixed fractional order ($\alpha=0.9$). 
  The cases represent healthy tissue ($D_f=1.5$), stage II-III malignant tumour ($D_f=1.7$), and theoretical saturation ($D_f=1.9$)~\cite{Elkington2022}. 
  (c) Global 2D mapping of the coagulation radius across the permissible parameter space ($D_f$, $d_s$) for a fixed $\alpha = 0.9$, showing isocontours.
  }
  \label{fig:topological_uncertainty}
\end{figure}

Having established the general behaviour of the coagulation zone under clinical constraints, we next quantify the absolute uncertainty introduced by the experimental indeterminacy of the spectral dimension $d_s$ (Fig.~\ref{fig:topological_uncertainty}). Consistent with previous observations, the absolute topological uncertainty of the coagulation radius increases with $\alpha$ (Fig.~\ref{fig:topological_uncertainty}a), exhibiting the same trend as the overall coagulation size itself. However, what deserves particular attention is that this absolute uncertainty is always highly significant in relative terms, meaning the margin of error is proportionally massive regardless of the diffusive regime. Focusing on the dependence of topological uncertainty on tumour evolution, \BLAU{while a clear trend cannot be extracted from Fig.~\ref{fig:topological_uncertainty}b, a wider exploration of the parameter space (Fig.~\ref{fig:topological_uncertainty}c) reveals a fundamental result. As evidenced by the distance between isocontours across the parameter space,} the sensitivity of the coagulation radius to variations of $d_s$ is highest in healthy tissue and decreases as the disease evolves (or as $D_f$ increases).

Considering that the possibility of topological variability among tissues is currently overlooked in the design of ablative protocols, the latter observation would materialise in clinical practice in two ways. On the one hand, the higher topological  sensitivity of the coagulation radius in the healthy parenchyma would imply that the primary driver of clinical variability is the surrounding healthy tissue and not the tumour itself. This aligns perfectly with recent clinical observations, suggesting that the physical properties of the surrounding healthy parenchyma—rather than the target tumour—are the main drivers of expansion of the ablation zone~\cite{Paolucci2022}. On the other hand, the theoretical reduction in uncertainty exhibited as the fractal dimension increases would clinically translate into a more predictable ablative response for tumours with a higher value of $D_f$. Indeed, \textit{in vivo} evidence shows that hepatocellular carcinoma exhibits a markedly lower standard deviation ($\pm$ 0.57~\si{\centi\meter}, corresponding to a coefficient of variation of $\sim$15\%, $n=17$) compared to colorectal ($\pm$ 0.75~\si{\centi\meter}, $\sim$20\%, $n=14$) or other liver metastases ($\pm$ 0.85~\si{\centi\meter}, $\sim$26\%, $n=12$)~\cite{Montgomery2004}. This is consistent with our findings, given that the fractal dimension of hepatocellular carcinoma (HCC) was found to be greater than that of secondary liver metastases (LM): $D_f^{HCC}=1.78$ and $D_f^{LM}=1.64$~\cite{Gheonea2014}.

\begin{figure}[htbp]
  \centering
  \includegraphics[width=\textwidth]{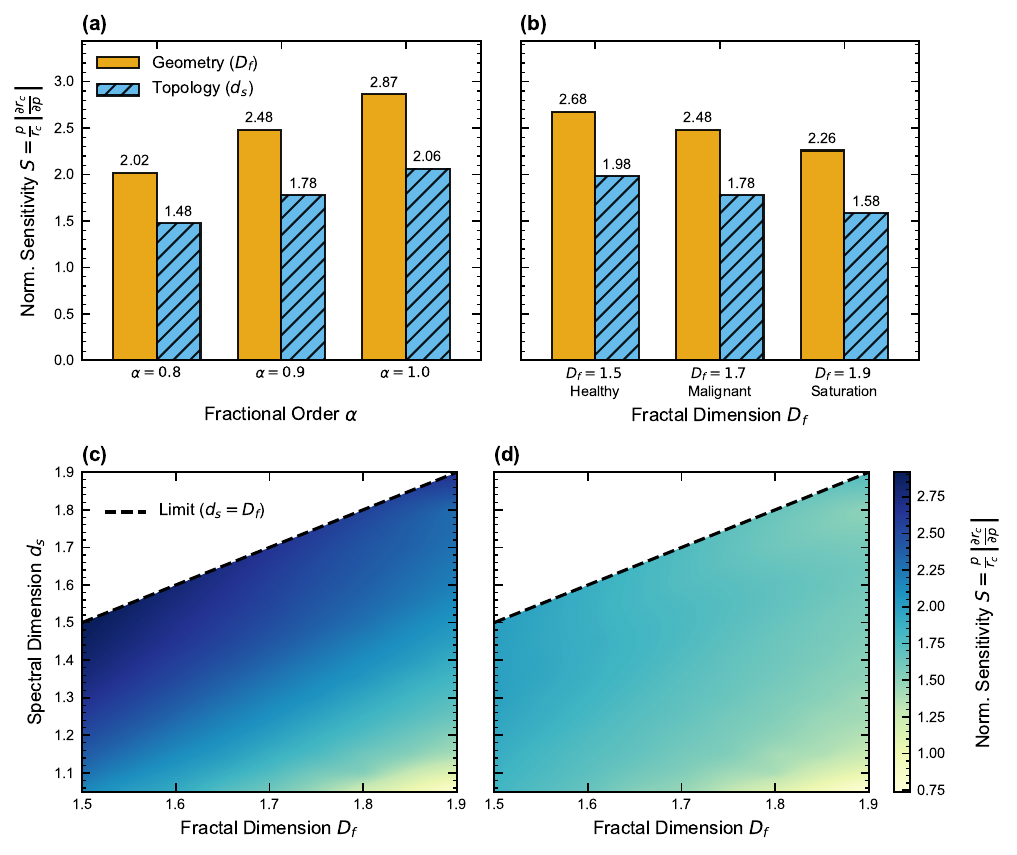}
  \caption{
  \textbf{Comparative local sensitivity of ablation size to geometric and topological variations.} 
  The normalised local sensitivity index ($S = (p/r_c)|\partial r_c/\partial p|$) is presented for the fractal dimension ($p=D_f$, geometry) and the spectral dimension ($p=d_s$, topology).
  (a) Sensitivity analysis across different fractional diffusive regimes ($\alpha$) for a fixed malignant tumour baseline ($D_f=1.7$, $d_s=1.5$).
  (b) Sensitivity analysis across three stages of tumour progression~\cite{Elkington2022} with fixed fractional order ($\alpha=0.9$): healthy tissue ($D_f=1.5$, $d_s=1.3$), stage II-III malignant tumour ($D_f=1.7$, $d_s=1.5$), and theoretical saturation ($D_f=1.9$, $d_s=1.7$). 
  (c, d) Global 2D heatmaps detailing the sensitivity to geometry (c) and topology (d) across the permissible $D_f$--$d_s$ parameter space for $\alpha=0.9$.
  }
  \label{fig:combined_sensitivity}
\end{figure}

It could be argued that the previous observations are solely a consequence of assuming that the spectral dimension of biological fractals spans across a broad range of values, especially considering that the actual variability could be more constrained. For this reason, we performed a local sensitivity analysis accounting for small perturbations ($\epsilon=5\%$) in both the spectral and fractal dimensions (Fig.~\ref{fig:combined_sensitivity}). While the sensitivity of the coagulation radius to geometry ($D_f$) is globally greater than the sensitivity to topology ($d_s$), the same fundamental trends found in the absolute uncertainty analysis are recovered here: the local sensitivity of the ablation radius to both $D_f$ and $d_s$ increases with $\alpha$ and diminishes as $D_f$ increases (as the tumour evolves). \BLAU{In addition, the global analysis (Fig.~\ref{fig:combined_sensitivity}c,d) reveals that the topological and geometrical sensitivities increase with $d_s$.} Notably, despite showing lower overall magnitudes than $D_f$, the sensitivity to topological changes remains significantly high across the studied domain. It reaches its maximum in the healthy tissue case and narrows the gap with the geometrical sensitivity as the fractional order $\alpha$ decreases. Finally, a sensitivity index predominantly greater than 1 ($S>1$) underscores the highly non-linear dependence of the coagulation radius on $D_f$ and $d_s$, \BLAU{with exceptions only appearing at the extreme limit of high $D_f$ and low $d_s$.} This emergent non-linearity is driven by the complex interplay between the space-dependent effective conductivity $\lambda_{\rm eff}(r)$, exponentially modulated by the fractal and spectral parameters, and the dynamic blood perfusion $\omega_b(T(r, t))$.

\subsection{Explaining clinical variability: topological rationale for the reduced efficacy in metastases}

To bridge the gap between theoretical modelling and clinical observation, we apply our framework to unravel a well-documented clinical anomaly: the reduced ablative efficacy in liver metastases compared to hepatocellular carcinoma~\cite{Heerink2018, Mathy2024}. To this end, we investigate whether the $\approx$15\% reduction in coagulation radius recently reported in clinical practice under identical ablative protocols (short-axis radius; $p=0.006$, $n=19$)~\cite{Mathy2024} can be explained by the distinct geometrical and topological nature of these tumours. For this analysis, the fractal dimensions of both types of tumours were fixed according to the literature~\cite{Gheonea2014}: $D_f^{HCC}=1.78$ and $D_f^{LM}=1.64$.

\begin{figure}[htbp]
  \centering
  \includegraphics[width=\textwidth]{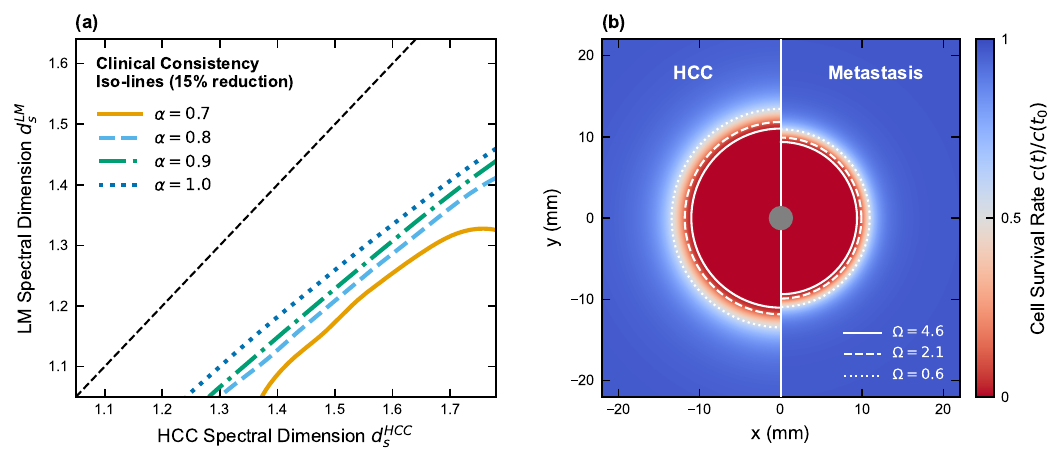}
  \caption{
  \textbf{Topological origin of clinical variability in tumour ablation.} 
  (a) Iso-contours represent the specific combinations of spectral dimensions ($d_s^{HCC}, d_s^{LM}$) required to reproduce the clinically observed $\approx 15\%$ 
  reduction in coagulation radius ($p=0.006$) for liver metastasis (LM) compared to hepatocellular carcinoma (HCC)~\cite{Mathy2024}. 
  Curves are plotted for varying fractional time-derivative orders $\alpha \in \{0.7, 0.8, 0.9, 1.0\}$, covering sub-diffusive to standard heat transfer regimes. 
  Fractal dimensions were fixed according to literature values: $D_f^{HCC}=1.78$ and $D_f^{LM}=1.64$~\cite{Gheonea2014}. 
  The region of clinical validity falls universally below the identity line (dashed), confirming $d_s^{LM} < d_s^{HCC}$.
  (b) Split heatmap illustrating the spatial distribution of cell survival rate ($c(t)/c(t_0)$) after 23~\si{\minute} of cooling, for the specific topological pair ($d_s^{HCC}=1.63$, $d_s^{LM}=1.36$) under standard diffusion 
  ($\alpha=1.0$). The left side depicts the primary tumour (HCC), while the right side shows the secondary tumour (Metastasis), 
  accurately reflecting the reported $\approx 15\%$ reduction in coagulation radius. White contours indicate ablation radii for coagulation 
  ($\Omega=4.6$, solid) and periablation ($\Omega=2.1$, dashed; $\Omega=0.6$, dotted).
  }
  \label{fig:HCCVsLM_alphas}
\end{figure}

Two key findings arise from this clinical application (Fig.~\ref{fig:HCCVsLM_alphas}a), both of which are universal across all $\alpha$-dependent diffusive regimes. First, there exists a region of validity such that the suitable combinations of $d_s$ lie strictly inside the theoretically admissible interval ($d_s \leq D_f \leq 2$). Second, and most importantly, this region is consistently located below the identity line ($d_s^{HCC} = d_s^{LM}$). As a direct consequence, not only can the fractal nature of the tissue explain the reported clinical variability (Fig.~\ref{fig:HCCVsLM_alphas}b), but the spectral dimension of liver metastases must also be lower than that of HCC ($d_s^{LM} < d_s^{HCC}$) for this differential ablative response to be reproduced. Physically, this implies that secondary tumours exhibit an intrinsically lower topological connectivity, which inherently restricts heat propagation and reduces the final ablation size.

Recent histopathological observations provide a compelling explanation for this phenomenon. 
Unlike primary hepatocellular carcinomas, liver metastases frequently exhibit a desmoplastic growth pattern characterised by a dense fibrotic rim~\cite{Latacz2022}, 
alongside a pathological accumulation of rigid extracellular matrix (ECM) components~\cite{Deng2025}. 
These dense fibrotic structures act as pronounced physical barriers~\cite{Deng2025} that disrupt the topological continuity of the tissue network 
($d_s^{LM} < d_s^{HCC}$), thereby hindering thermal diffusion and fundamentally explaining the reduced ablation efficacy consistently observed in clinical practice.

Although the proposed framework provides a robust topological basis for the clinical variability observed in thermal ablation, 
certain limitations of the current study must be acknowledged. 
First, assuming a spatially uniform fractal dimension ($D_f$), our computational domain is effectively restricted to the targeted tumour, 
neglecting the distinct structural properties of the surrounding healthy parenchyma. 
To maximise the accuracy of future predictive models, a multi-domain or spatially heterogeneous approach (i.e., a spatially varying $D_f$) should be adopted. 
Additionally, more research is required to rigorously define the spatial boundaries of the fractal self-similarity regime, 
which in this study was assumed inherently to span continuously from the microscopic scale at which $D_f$ is measured in the literature.

\section{Conclusions}
\label{sec:conclusions}

In this work, we have established that the efficacy of clinical thermal ablation is fundamentally based on the anomalous heat transport dynamics inherent to the fractal nature of the tumour. 
By modelling the closed-loop power control routinely used in clinical settings, we revealed that the spatial expansion of the coagulation zone is deeply governed by the fractal geometry of the tissue 
($D_f$) and, critically, its topological connectivity ($d_s$). Furthermore, we identified topological indeterminacy as a primary driver of clinical unpredictability, 
which we showed to strictly decrease alongside tumour evolution and structural complexity. 
Consequently, this framework provides robust physical insight into two key clinical observations: 
the dominant role of the surrounding healthy parenchyma in dictating final ablation outcomes, 
and the significantly higher standard deviation in ablation zone size reported for tumours with lower fractal dimensionality.

In particular, we shed light on the physical interpretation of the spectral dimension in biological statistical fractals while successfully unravelling a clinical anomaly: 
significantly reduced ablative efficacy in liver metastases compared to primary hepatocellular carcinomas. 
Our model translates this clinical variability into a fundamental topological inequality ($d_s^{LM} < d_s^{HCC}$), 
linking the topological connectivity dictated by $d_s$ with desmoplastic growth and dense fibrotic barriers characteristic of metastatic tumours.

The robustness and universal validity of our findings were further demonstrated by studying them across a wide range of fractional orders $\alpha$, 
effectively accounting for 
non-Markovian effects across \BLAU{the physically relevant} 
sub-diffusive to standard 
\BLAU{heat transfer} regimes.

Looking ahead, while direct clinical implementation remains a future milestone, 
this fractal biothermal framework offers a promising avenue to optimise treatment planning, improve ablation efficacy, and minimise local recurrence. 
However, the paradigm shift towards fractal-aware thermal dosimetry requires further research into the experimental determination of the spectral dimension $d_s$, 
potentially through the inverse application of this model. Beyond tumour ablation, our mathematical approach holds 
tremendous translational potential for other physical and biological systems governed by anomalous diffusion in heterogeneous fractal media.

\section*{Acknowledgments}
M.O.-F. acknowledges support from the Erasmus+ Programme of the European Union (Project No. 2024-1-ES01-KA131-HED-000211014), 
which enabled an academic mobility at Universit\'e de Lorraine (France), during which this work was initiated.

\clearpage
\appendix
\section{Parametric robustness and sensitivity analysis}
\label{sec:appendix_sensitivity}
\setcounter{figure}{0} 
\renewcommand{\thefigure}{A\arabic{figure}}
As described in Section~\ref{sec:model}, a local sensitivity analysis was performed to evaluate the parametric robustness of the model. 
The detailed behaviour of the sensitivity $S$ (Eq.~\ref{eq:sensitivity}) of the coagulation radius with respect to $D_f$ and $d_s$ is illustrated in 
Figure~\ref{fig:sensitivity_panel}, under variations of the characteristic length, relaxation time, and limit temperature.
\begin{figure}[H]
  \centering
  \includegraphics[width=\textwidth]{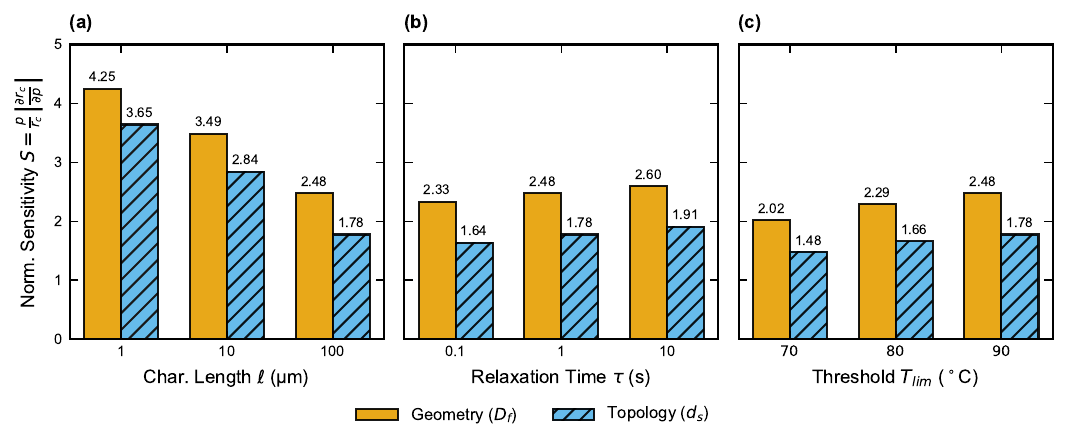}
  \caption{\textbf{Local sensitivity analysis and parametric robustness.} 
  The normalised sensitivity index $S$ of the coagulation radius was computed locally relative to the baseline set of values \{$D_f=1.7$, $d_s=1.5$\} with a fixed fractional order $\alpha=0.9$. (a) Characteristic Length ($\ell$): 
  The sensitivity increases as the characteristic length of the fractal decreases, showing a reduction in the gap between the geometrical and topological sensitivities. 
  (b) Relaxation Time ($\tau$): The sensitivity hierarchy is robust and insensitive to relaxation time variations; geometry ($D_f$) consistently outweighs topology ($d_s$) 
  across the entire $\tau$ range, maintaining a nearly constant dominance ratio, despite a slight overall increase in sensitivity. 
  (c) Limit Temperature ($T_{\rm lim}$): Increasing the target temperature amplifies the system's absolute sensitivity 
  (indicating higher uncertainty at higher temperatures) but preserves the strict predominance of geometric effects over topological ones.}
  \label{fig:sensitivity_panel}
\end{figure}

\clearpage
\printbibliography

\end{document}